\documentclass[12pt]{iopart}
\usepackage{iopams}
\usepackage{xcolor}
\usepackage[breaklinks=true,colorlinks=true,linkcolor=blue,urlcolor=blue,citecolor=blue]{hyperref}
\expandafter\let\csname equation*\endcsname\relax
\expandafter\let\csname endequation*\endcsname\relax
\usepackage{amsmath}
\usepackage{footmisc}

\allowdisplaybreaks

\begin{document}

\title[Extending the symmetry of the massless Klein-Gordon equation]{Extending the symmetry of the massless Klein-Gordon equation under the general disformal transformation}

\author{Allan L. Alinea\footnote{corresponding author} and Mark Ruel D. Chua}

\address{Institute of Mathematical Sciences and Physics, University of the Philippines Los Ba\~nos, 4031 College, Los Ba\~nos, Laguna, Philippines}
\ead{alalinea@up.edu.ph}
\vspace{10pt}
\begin{indented}
\item[]August 2022
\end{indented}

\begin{abstract}
The Klein-Gordon equation, one of the most fundamental equations in field theory, is known to be \textit{not} invariant under conformal transformation. However, its massless limit exhibits symmetry under Bekenstein's disformal transformation, subject to some conditions on the disformal part of the metric variation. In this study, we explore the symmetry of the Klein-Gordon equation under the general disformal transformation encompassing that of Bekenstein and a hierarchy of `sub-generalisations' explored in the literature (within the context of inflationary cosmology and scalar-tensor theories). We find that the symmetry in the massless limit can be extended under this generalisation provided that the disformal factors takes a special form in relation to the conformal factor. Upon settling the effective extension of symmetry, we investigate the invertibility of the general disformal transformation to avoid propagating non-physical degrees of freedom upon changing the metric. We derive the inverse transformation and the accompanying restrictions that make this inverse possible.
\end{abstract}


\section{Introduction}
\textit{Disformal transformation} \cite{Bekenstein:1992pj} is a metric transformation introduced by Bekenstein in 1992 as a generalisation of the more commonly known \textit{conformal transformation} \cite{Wald:1984rg,Groen:2007zz}. The aim was to relate Finsler and Riemann geometries in a single gravitational theory. Since then, disformal transformation found its utility in many cutting-edge research areas in Physics beyond the original scope of Bekenstein's work. In the study of scalar-tensor theories, disformal transformation plays an important role in relating one action to another. This is with an insight to solve pressing problems about the Universe such as dark energy and dark matter, and investigate the physical consistency of the resulting action upon transformation; see Refs.~\cite{Kobayashi:2019hrl,Gleyzes:2014qga,Bettoni:2013diz,Sakstein:2014isa,Langlois:2015cwa,Achour:2016rkg,Alinea:2020sei}. In the area of inflationary cosmology, disformal transformation is used in the study of G-inflation and beyond \cite{Kobayashi:2011nu,Horndeski:1974wa}, Higgs inflation {\cite{Sato:2017qau}}, and disformal inflation {\cite{Kaloper:2003yf}}, to name a few. The investigation of disformal transformation also extends to hairy black holes \cite{BenAchour:2019fdf}, D-brane cosmology \cite{Koivisto:2013fta}, particle motion in semi-classical gravity \cite{Chowdhury:2021fjw}, etc.

As in any other areas of physics involving the study of transformation, symmetry is an `omnipresent' idea that has to be taken into account. Of particular interest are transformations that lend physical observables and equations of motion describing physical systems, invariant. For disformal transformation, it is found that both the gauge-invariant scalar and tensor primordial cosmological perturbations \cite{Dodelson:2003ft,Mukhanov:2005sc,Alinea:2015pza,Alinea:2016qlf,Alinea:2017ncx}---the seeds that gave rise to the current universe---are invariant under a hierarchy of this transformation \cite{Alinea:2020laa, Minamitsuji:2014waa, Tsujikawa:2014uza,Domenech:2015hka,Motohashi:2015pra}. Furthermore, the `trinity' of relativistic field equations, namely, Klein-Gordon equation \cite{Falciano:2011rf}, Maxwell's equations \cite{Goulart:2013laa}, and Dirac equation (under the Inomata condition) \cite{Bittencourt:2015ypa} are symmetric with respect to disformal transformation, subject to some constraints on the conformal and disformal factors in the variation.

The focus of this work is the disformal transformation of the Klein-Gordon equation. Falciano and Goulart \cite{Falciano:2011rf} showed that subject to some conditions, the massless Klein-Gordon equation is invariant under the disformal transformation of the metric $g_{\mu\nu}$ that we rewrite as\footnote{Ref.~\cite{Falciano:2011rf} started out with what we identify as the disformal transformation of the inverse metric. Here, we follow a slightly different notation; i.e., $A_{FG} = 1/A$ and $B_{FG} = -B/[A(A - 2BX)]$, where the subscript `FG' pertains to Ref.~\cite{Falciano:2011rf}.} 
\begin{align}
	\label{usualDT}
	g_{\mu\nu}
	\rightarrow
	\widehat g_{\mu\nu}
	=
	A(\phi, X)g_{\mu\nu} + B(\phi, X) \phi_{;\mu}\phi_{;\nu},
\end{align}
where $A$ and $B$ are the conformal and disformal factors, respectively. They depend on the scalar field $\phi$ and the kinetic term $X$ defined as $ X \equiv -\tfrac{1}{2} g^{\mu\nu} \phi_{;\mu}\phi_{;\nu}$. Subject to the condition given by\footnote{The constant $c$ is a minor correction in the result in Ref.~\cite{Falciano:2011rf} where $c^2$ is simply unity.}
\begin{align} 
	\label{dtcondn}
	B = \frac{A - c^2A^{n-1}}{2X},
	\quad
	(c = \text{const.})
\end{align}
where $n$ is the dimension of the background spacetime, the Klein-Gordon equation for the scalar field with mass $m$, transforms\footnote{We are using the metric signature $(-,+,+,+)$.} as $ \square \phi - m^2 \phi = 0 \rightarrow (\square \phi)/(c^2A^{n-1}) - m^2\phi = 0 $. Consequently, it is invariant under the disformal transformation in the massless limit. For the massive case, the mass may be seen as transforming from $m^2 \rightarrow m^2/(c^2 A^{n-1})$.

The transformation given by (\ref{usualDT}) is not the only form of disformal transformation. It can be generalised to a form involving higher-order derivatives of $\phi$ and/or derivatives of the kinetic term. It is then interesting to see if the Klein-Gordon equation would remain invariant under such generalised transformation, and identify the needed conditions, if any, for the preservation of symmetry. In this work, we consider a general disformal transformation given by\footnote{As in Bekenstein's disformal transformation given by (\ref{usualDT}), this generalisation does not necessarily preserve the metric signature. Interested readers may refer to Ref.~\cite{Bettoni:2013diz} and references therein for the discussion on restrictions involving Lorentzian signature and causality. } \cite{Alinea:2020laa,Takahashi:2021ttd}
\begin{align}
	\label{genDT}
	g_{\mu\nu}
	\rightarrow
	\widehat g_{\mu\nu}
	=
	Ag_{\mu\nu}
	+
	\Phi_\mu \Phi_\nu,
	\quad\text{where}\quad
	\Phi_\mu 
	\equiv 
	C\phi_{;\mu}
	+
	DX_{;\mu},
\end{align}
with $A, C,$ and $D$ being functionals of $(\phi, X,Y,Z)$, in general, and the quantities $Y$ and $Z$ are defined as $ Y \equiv g^{\mu\nu}\phi_{;\mu}X_{;\nu}$ and $ Z \equiv g^{\mu\nu}X_{;\mu}X_{;\nu}$. Note that this transformation involves a derivative of the kinetic term in addition to that of the scalar field; effectively, the second-order derivative of $\phi$ is encoded in $X_{;\mu}$. It reduces to the special case of Bekenstein given by (\ref{usualDT}) when $D = 0$ and the dependencies on $(Y,Z)$ are removed; in this case, we can identify $C^2(\phi,X) = B(\phi,X)$ in (\ref{usualDT}). 

The form of the disformal transformation given by (\ref{genDT}) is a generalisation of the hierarchy of transformations investigated in Ref.~\cite{Minamitsuji:2014waa,Tsujikawa:2014uza,Domenech:2015hka,Motohashi:2015pra} within the context of inflationary cosmology (see Refs.~\cite{Firouzjahi:2018xob,Domenech:2018vqj} for other forms covering dual scalars and vector disformal transformation.)  Moreover, it is inspired by Refs.~\cite{Alinea:2020laa} and \cite{Takahashi:2021ttd}. In Ref.~\cite{Alinea:2020laa}, it is shown that both the scalar and tensor primordial cosmological cosmological perturbations are invariant under (\ref{genDT}) in the superhorizon limit \cite{Dodelson:2003ft,Mukhanov:2005sc}.  On the other hand, it is explored in Ref.~\cite{Takahashi:2021ttd} within the scope of invertibility and physical degrees of freedom in scalar-tensor theories. 

The first study finds its relevance in connecting general actions through disformal transformation and showing the physical quantities that are symmetric. This is much like moving to and from the Einstein and Jordan frames in conformal transformation while retaining the character of primordial cosmological perturbations. The second study complements the first one in view of the fact that metric transformation may lead to Ostrogradsky instability \cite{Ostrogradsky:1850fid,Woodard:2015zca} and propagating non-physical degrees of freedom \cite{Domenech:2015tca,Takahashi:2017zgr,Gleyzes:2014dya,Zumalacarregui:2013pma,Crisostomi:2016czh,BenAchour:2016fzp} in scalar-tensor actions in which the Klein-Gordon action belongs; invertibility ensures that the number of propagating physical degrees of freedom remain the same under a metric transformation. Having been explored in Refs. \cite{Alinea:2020laa} and \cite{Takahashi:2021ttd} the `umbrella' disformal transformation encompassing that of Bekenstein and those of other works \cite{Minamitsuji:2014waa,Tsujikawa:2014uza,Domenech:2015hka,Motohashi:2015pra}, it is time that we revisit one of the most, if not the most fundamental equation in field theory, namely, the Klein-Gordon equation, under the general disformal transformation.

To be clear, our aim in this study is two-fold. The first one is to determine the form of the disformal factors $ C $ and $ D $ such that the massless Klein-Gordon equation is invariant under the general disformal transformation given by (\ref{genDT}); this is in the same spirit as in Ref.~\cite{Goulart:2013laa}. The second one is to demonstrate the invertibility of the general disformal transformation subject to the result of the first one and other auxiliary conditions. In particular, we wish to write $ g_{\mu \nu } $ in terms of disformally transformed quantities to the extent permitted by the form of the conformal and disformal factors in the transformation.

To meet these ends, this paper is organised as follows. In Sec. {\ref{secDTKG}}, we derive the general disformal transformation of the d'Alembertian of the scalar field, $\square \phi$, following a pathway to symmetry of the Klein-Gordon equation. Along the way, we determine the conditions on the functionals $C$ and $D$,in relation to the conformal factor $ A $, that render the massless Klein-Gordon equation invariant. In addition to this, we also look at the transformation of the mass in the massive case. In Sec. {\ref{secDTKGshort}}, we provide an alternative way to derive the conditions in the previous section for the symmetry; this serves to confirm our results. In Sec. \ref{secDTinvert}, we comment on and derive the inverse general disformal transformation covered in this work. Lastly, we state our concluding remarks and future prospects in Sec. \ref{secConclud}.

\section{Symmetry and the General Disformal
Transformation of the Klein-Gordon Equation}
\label{secDTKG}
In this section, we wish to derive the disformal transformation (\ref{genDT}) of the d'Alembertian of the scalar field contained in the Klein Gordon equation. The original background $ n $-dimensional spacetime is described by the metric, $ g_{\mu \nu } $, with signature $ (-,+,+,+) $. We assume, as in Ref.~\cite{Bettoni:2013diz}, that the transformation leading to $ \widehat g_{\mu \nu } $, preserves the Lorentzian signature and causality structure of spacetime. Cases where the disformal and conformal factors lead to flipping of Lorentzian signature and acausal behaviour are left for future studies. 

In so finding the general disformal transformation of the Klein-Gordon equation, our pathway is bent in search of symmetry for the massless case. In other words, we do not simply derive the full transformation of the equation under (\ref{genDT}) but take opportunities along the way whenever they present themselves as conditions that we may impose to have an invariant Klein-Gordon equation.

The d'Alembertian of the scalar field, $\square \phi $, changes under the disformal transformation to $\widehat \square \phi = \widehat g^{\mu\nu}\widehat \nabla_\mu\widehat \nabla_\nu \phi$, where $\widehat g^{\mu\nu}$ is the inverse of the disformal metric given by (\ref{genDT}) above. The hatted covariant derivative operators (as in all other hatted quantities in this work) are disformally transformed quantities. From the \textit{Sherman-Morrison formula} for matrix inversion, we find
\begin{align}
	\label{genDTinv}
	\widehat g^{\mu \nu }
	&=
	\frac{g^{\mu \nu }}{A}
	-
	\frac{\Phi ^{\mu} \Phi ^{\nu} }{A(A - 2\chi)},
	\quad\text{where}\quad
	\chi 
	\equiv 
	-\tfrac{1}{2}g^{\mu \nu }\Phi _\mu \Phi _\nu.
\end{align}
It is straightforward to show that $\widehat g^{\mu\alpha}\widehat g_{\alpha\nu} = \delta^\mu_\nu$, verifying the validity of the expression for the inverse disformal metric above.

For the covariant derivatives of the scalar field, we have $\widehat \nabla _\mu \widehat \nabla _\nu \phi = \phi _{;\nu ,\mu } - \widehat \Gamma ^\alpha  _{\mu \nu }\phi _{;\alpha },$ where the symbols comma `,' and semicolon `;' mean partial differentiation and covariant differentiation, respectively\footnote{Needless to say, for the scalar field, $\phi_{;\alpha} = \phi_{,\alpha}$, but $\phi_{;\alpha\beta} \ne \phi_{,\alpha\beta}$.}. The hatted Christoffel symbol on the right hand side can be expressed in terms of the disformal metric, derivative thereof, and its inverse, as $ \widehat \Gamma ^\alpha _{\mu \nu } = \tfrac{1}{2} \widehat g^{\alpha \beta }\left(		\widehat g_{\beta \mu ,\nu } + \widehat g_{\nu \beta ,\mu }	- \widehat g_{\mu \nu ,\beta }\right)$. Using the equations for the disformal metric and its inverse given by (\ref{genDT}) and (\ref{genDTinv}), respectively, the hatted Christoffel symbol can be decomposed as a sum of the original Christoffel symbol and a tensor term. In particular,
\begin{align}
	\label{chtoffhat}
	\widehat \Gamma ^\alpha _{\mu \nu }
	&=
	\Gamma ^\alpha _{\mu \nu }
	+
	C^\alpha _{\mu \nu },
\end{align}
where $C^\alpha _{\mu \nu }$, the contribution due to the disformal part of the transformation, is given by 
\begin{align}
	\label{Calphamunu}
	C^\alpha _{\mu \nu }
	&=
	\frac{
		A_{;\nu }\delta ^\alpha _\mu 
		+
		A_{;\mu }\delta ^\alpha _\nu 
		-
		A^{;\alpha }g_{\mu \nu }	
	}{2A }
	-
	\frac{
		\Phi ^\alpha(
		A _{;\nu }\Phi _\mu 
		+
		A _{;\mu }\Phi _\nu 
		-
		A _{;\beta }\Phi ^\beta g_{\mu \nu }
		)
	}{2A (A - 2\chi )} 
	\nonumber
	\\[0.5em]
	&\qquad
	+\,
	\left[
	\frac{g^{\alpha \beta }}{A}
	-
	\frac{\Phi ^\alpha \Phi ^\beta }{A(A - 2\chi )} 
	\right]
	(
	\Phi _\mu\Phi _{[\beta;\nu ]} 
	+
	\Phi _{\nu }\Phi _{[\beta;\mu]} 	
	)	
	+
	\frac{\Phi^\alpha\Phi _{(\mu;\nu )}}{A - 2\chi }.    
\end{align}
Here, $\Phi_{[\mu;\nu]} \equiv \tfrac{1}{2}(\Phi_{\mu;\nu} - \Phi_{\nu;\mu})$ and $\Phi_{(\mu;\nu)} \equiv \tfrac{1}{2}(\Phi_{\mu;\nu} + \Phi_{\nu;\mu})$. Observe that $C^\alpha_{\mu\nu}$ is symmetric with respect to the lower indices. It follows that $\widehat \Gamma ^\alpha_{\mu\nu}$ retains the symmetry of the original $\Gamma^\alpha_{\mu\nu}$. In the limit where $C = D = 0$, it reduces to that of the familiar conformal transformation \cite{Wald:1984rg}; i.e., $\widehat \Gamma^\alpha_{\mu\nu} - \Gamma^\alpha_{\mu\nu}$ $= C^\alpha_{\mu\nu} \rightarrow (A_{;\nu }\delta ^\alpha _\mu + A_{;\mu }\delta ^\alpha _\nu -	A^{;\alpha }g_{\mu \nu })/2A$.

All the needed expressions are now in place to solve for $\widehat \square \phi$. We have upon using (\ref{chtoffhat}) for the hatted Christoffel symbol in the expression for the d'Alembertian,
\begin{align}
	\label{sqdAlem}
	\widehat \square \phi	
	&=
	\frac{\square\phi }{A}
	-
	\frac{g^{\mu \nu} C^\alpha _{\mu \nu }\phi _{;\alpha }}{A}  
	-
	\frac{\Phi ^{\mu} \Phi ^{\nu} \phi _{;\mu \nu }}{A(A - 2\chi)} 
	+
	\frac{\Phi ^{\mu} \Phi ^{\nu} C^\alpha _{\mu \nu }\phi _{;\alpha }}{A(A - 2\chi)}.
\end{align}
We see that the transformed d'Alembertian of $\phi$ is a sum of the original d'Alembertian of $\phi$, scaled by the conformal factor, $A$, and terms due to the disformal part of the transformation. While these terms are straightforward to compute using the equation for $C^{\alpha}_{\mu\nu}$ given above, the result is quite long and the calculation is laborious. With an eye for symmetry, as alluded above, we opt to cancel terms early on by imposing `symmetry' condition(s) on the disformal factors $ C $ and $ D $, leading to the invariance of the massless Klein-Gordon equation.

This cancellation starts with the second term on the right hand side of the equation above for $\widehat \square \phi$. It can be expressed as 
\begin{align*}
	g^{\mu \nu }C^\alpha _{\mu \nu }\phi _{;\alpha }
	&=
	\frac{
		(2-n)\phi ^{;\alpha }A_{;\alpha }
	}{2A }
	-
	\frac{
		(2-n)\phi _{;\alpha }\Phi ^\alpha\,
		A _{;\beta  }\Phi ^\beta 
	}{2A (A - 2\chi )}
	+
	\frac{2\phi ^{;\alpha  }\Phi ^\beta \Phi _{[\alpha ;\beta  ]}}{A}
	+
	\frac{\phi _{;\alpha }\Phi^\alpha \Phi^\beta {}_{;\beta  }}{A - 2\chi }.  
\end{align*}
Observe that of all the terms in the equation for $\widehat \square  \phi $ above, only this term generates $ \square X $. Furthermore, only the term involving $ {\Phi ^\beta }_{;\beta } $ on the right hand side of the equation above for $\phi _{;\alpha }g^{\mu \nu }C^\alpha _{\mu \nu }$ generates $ \square X $; to wit, $ {\Phi ^\beta }_{;\beta } = C_{;\beta }\phi ^{;\beta } + C\square \phi + D_{;\beta }X^{;\beta } + D\square X$. If $ \widehat \square \phi  $ is then to be proportional to $ \square \phi $, leading to the equality between the two in the massless limit, the term involving $ \phi _{;\alpha }\Phi^\alpha \Phi^\beta {}_{;\beta  } $ above should vanish. For this to happen, we may impose the ``orthogonality'' condition\footnote{With $ \Phi^\alpha = C\phi ^{;\alpha } + DX^{;\alpha }$, they are not orthogonal in the usual sense of the word orthogonal in Linear Algebra and Hilbert spaces; hence, the quotation marks.} between $ \Phi ^\alpha  $ and $ \phi _{;\alpha } $; that is, 
\begin{align}
	\label{ortphi}
	\Phi ^\alpha  \phi _{;\alpha } = 0.
\end{align}
As a consequence of this condition, we have the functional relation given by $ DY = 2CX $. At this point, this relation can be satisfied by infinitely many combinations of $ D(\phi ,X,Y,Z) $ and $ C(\phi ,X,Y,Z) $; e.g., $ C = Y $ and $ D = 2X $. However, as we shall see below, $ C $ and $ D $ should take special forms in relation to $ A $ and $ (X,Y,Z) $ for the symmetry of the massless Klein-Gordon equation.

Continuing, we find that given the ``orthogonality'' condition, the substitution of the equation for $C^\alpha_{\mu\nu}$ given by (\ref{Calphamunu}) in the equation for $\widehat \square \phi$ given by (\ref{sqdAlem}) leads us to a much simpler expression for the latter.
\begin{align}
	\label{sqdAlem2}
	\widehat \square \phi 
	&=
	\frac{\square\phi }{A}
	+
	\frac{[n(A - 2\chi) -2A + 6\chi ]\phi ^{;\alpha }A_{;\alpha }}
	{2A^2(A - 2\chi)}
	-
	\frac{
		\Phi ^{\alpha}\phi _{;\alpha \beta }\Phi ^{\beta} 
		+
		2\phi ^{;\alpha }
		\Phi ^\beta \Phi _{[\alpha ;\beta  ]}
	}{A(A - 2\chi)}.
\end{align}
Needless to say, if the conformal factor is unity and $C=D=0$, then $\widehat \square \phi = \square \phi$. On the other hand, when the transformation is purely conformal, we find both $ \chi$ and $\Phi _\mu$ vanish, and gain the expected result given by\footnote{We cannot set $ D = 0 $ independent of $ C $ as a consistency check because the condition $ \phi ^{;\alpha }\Phi _{;\alpha } = 0 $ imposes $ D\propto C $. Setting $ D = 0 $ also means $ C = 0 $ or $ B = 0 $ (since $ C^2 = B $).}
\begin{align}
	\widehat \square \phi 
	&=
	\frac{\square\phi }{A}
	+
	\frac{(n-2)\phi ^{;\alpha }A_{;\alpha }}{2A^2}.
\end{align}
Interestingly, the Klein-Gordon equation is \textit{not} symmetric with respect to conformal transformation unless the dimension of the embedding manifold is $n=2$. Being able to push it to symmetry \cite{Falciano:2011rf} elevates the value of Bekenstein's disformal transformation. In this work, we push the symmetry further under the general disformal transformation (\ref{genDT}) and uncover the conditions for this symmetry.

Beyond the conformal limit, the factors $C$ and $D$ are not necessarily vanishing but are bound by the ``orthogonality'' condition given by (\ref{ortphi}). As a consequence, for the numerator of the last term on the right hand side of (\ref{sqdAlem2}), we have
\begin{align}
	\label{sqdAlem23rdterm}
	\Phi ^\mu\phi _{;\mu \nu } \Phi ^\nu 
	+
	2\phi ^{;\alpha  }	\Phi ^\beta \Phi _{[\alpha ;\beta  ]} 
	&=
	-\frac{B}{Y} (Y^2 + 4XZ)
	-
	\frac{X}{Y^2} (Y^2 + 2XZ)\phi ^{;\alpha }B_{;\alpha }	
	\nonumber
	\\[0.5em]
	&\qquad
	-\,
	\frac{2B X^2\phi ^{;\alpha }Z_{;\alpha}}{Y^2}		
	+
	\frac{4BX^2Z\phi ^{;\alpha }Y_{;\alpha }}{Y^3},  
\end{align}
where we identify $C^2 = B$ from the correspondence of the usual disformal transformation (\ref{usualDT}) and the general disformal transformation (\ref{genDT}).

If $\widehat \square \phi$ is to be proportional to $\square \phi$ in (\ref{sqdAlem2}), then the two terms following the first term involving $\square\phi$ on the right hand side should vanish. For this to happen, $B$ may be related to $A$. In an attempt to find this relationship, we take the \textit{ansatz} $B = q(X,Y,Z) h(X) f(A)$, where the functionals $q, h,$ and $f$ are to be determined. Note that in this \textit{ansatz}, the dependence on $\phi$ is encoded in $A$ alone while $X$ is distributed amongst $q, h,$ and $A$, with $q$ (and $A$) being also dependent on $(Y,Z)$. Using this \textit{ansatz} and (\ref{ortphi}) we find
\begin{align}
	q &= \frac{Y^2}{Y^2 + 2XZ},
	\quad
	h
	=
	\frac{1}{X},
	\quad\text{and} \quad
	f
	=
	-\chi.
\end{align}

We conclude that under the disformal transformation given by $\widehat g_{\mu\nu} = Ag_{\mu\nu} + \Phi_{\mu}\Phi_{\nu}$, the Klein-Gordon equation for scalar field $\phi$ transforms as
\begin{align}
	\boxed{
		\square \phi - m^2\phi = 0
		\quad\rightarrow\quad
		\frac{\square \phi}{A} - m^2\phi = 0,
	}
\end{align}
provided that the disformal factors $ C $ and $ D $ are related to the conformal factor and $ (X,Y,Z) $ as
\begin{align}
	\label{condsymkg}
	\boxed{
		C^2
		=
		\frac{Y^2(bA^{3 - n}-A)}{2X(Y^2 + 2XZ)}
		\quad\text{and}\quad
		D^2
		=
		\frac{2X(bA^{3 - n}-A)}{Y^2 + 2XZ}.
	}
\end{align}

\noindent
In the massless limit, the Klein-Gordon equation is invariant under the transformation. On the other hand, for the massive case, the mass may be seen as transforming by conformal rescaling (i.e., $m^2\rightarrow m^2/A$) from the manifold described by $g_{\mu\nu}$ to the manifold described by $\widehat g_{\mu\nu}$. 

It is worth noting that our result here for the massive field is a little different from that of the usual disformal transformation, $\widehat g_{\mu\nu} = Ag_{\mu\nu} + B\phi_{;\mu}\phi_{;\nu}$, wherein $ \square \phi - m^2 \phi = 0 \rightarrow (\square \phi)/(c^2A^{n-1}) - m^2\phi = 0 $, leading to $m^2 \rightarrow m^2/(c^2 A^{n-1})$ \cite{Falciano:2011rf}. The dependence on the dimension $n$ is apparent in the usual one, while for our transformation, it is not present. In fact, for the massive field, the transformation of the Klein-Gordon equation is not simply a special case of ours. The condition that relates $B$ to $A$ given by (\ref{dtcondn}) for the usual disformal transformation is not a special case of the second of (\ref{condsymkg}) due to the ``orthogonality'' condition binding $C$ and $D$. This leads to different mass transformations with one depending on $n$ while the other one does not. However, the massless limit yields the same symmetry for both cases. It is in this sense that we are extending the symmetry of the massless Klein-Gordon equation under the general disformal transformation (\ref{genDT}).

Before we leave this section, it is good to clarify one point leading to (\ref{condsymkg}) and consequently, to the invariance of the Klein-Gordon equation. Along the way of our calculation, we encountered the `orthogonality' condition, namely, $\Phi^\alpha\phi_{;\alpha} = 0$. Taken at face value, this seems to imply a relation for $ \phi $ in addition to the Klein-Gordon equation, making it possibly over determined. However, a better and correct way to look at the `orthogonality' condition is as part of the intermediate steps leading to (\ref{condsymkg}). In view of the idea that this result should not be taken as an equation for $ \phi $ but the forms $C$ and $D$ should take in order for the equation given by $ \square\phi = 0 $, to be invariant, the `orthogonality' relation, $ \Phi^\alpha\phi_{;\alpha} = 0 \,\Leftrightarrow\,2CX = DY $, should be seen as a functional relationship between $ C $ and $ D $ and a stepping stone leading to their desired forms given by (\ref{condsymkg}). Taking our derivation in this section as a whole, it may not be taken as an auxiliary equation for $ \phi $.

\section{Alternative Derivation Leading to the Conditions for Symmetry}
\label{secDTKGshort}
The immediately preceding section saw the calculation of $\widehat \square \phi$ using a direct `expansion' involving the Christoffel symbol. In this section, we present a slightly different pathway, used in Ref.~\cite{Falciano:2011rf} for the usual disformal transformation, to the conditions leading for symmetry of the massless Klein-Gordon equation under the general disformal transformation given by (\ref{genDT}). The process of calculation is mainly for verification of the previous section's results.

Following this pathway, we start from the relation involving double covariant differentiation, namely, $	\square \phi = g^{\mu \nu }\phi _{,\nu \mu } - g^{\mu \nu }\Gamma ^{\alpha }_{\mu \nu }\phi _{;\alpha }$. Knowing, however, that the contraction of the metric and the Christoffel symbol can be written as $g^{\mu \nu }\Gamma ^{\alpha }_{\mu \nu } = -(g^{\alpha \beta }\sqrt{-g} )_{,\beta }/\sqrt{-g} $, leads us to a compact form given by 
\begin{align}
	\label{dAlemshort2}
	\square \phi 
	&=
	\frac{(g^{\mu \nu }\sqrt{-g} \phi _{,\mu })_{,\nu}}{\sqrt{-g} }.
\end{align}
Because the derivative operators involved on the right hand side are not covariant derivative operators but only partial derivative operators, the disformal transformation of $ \square \phi  $ can simply be written as
\begin{align}
	\label{dAlemshort}
	\widehat \square \phi 
	&=
	\frac{(\widehat g^{\mu \nu }\,\sqrt{-\widehat g}\,
		\phi _{,\mu })_{,\nu}}{\sqrt{-\widehat g} }.
\end{align}

With the inverse disformal metric already given by (\ref{genDTinv}), we only need to find the hatted determinant. Given the form of the disformal metric provided by (\ref{genDT}), we find from the \textit{matrix determinant lemma}, $\widehat g = A^{n-1}g(A - 2\chi )$, where $\chi$ is defined by the second of (\ref{genDTinv}). Using this equation for $\widehat g$ and (\ref{genDTinv}) for $\widehat g^{\mu\nu}$, in the sub-expression $\widehat g^{\mu \nu }\,\sqrt{-\widehat g}\,\phi _{,\mu }$ above contained in the equation for $\widehat \square \phi$, yields
\begin{align}
	\label{ggphi}
	\widehat g^{\mu \nu }\sqrt{-\widehat g}\, 
	\phi _{,\mu }
	&=
	A^{\frac{n-3}{2} }(A - 2\chi )^{\frac{1}{2} }
	g^{\mu\nu}\sqrt{-g}\,\phi_{;\mu } 
	- 
	\frac{A^{\frac{n-3}{2} }\sqrt{-g} }{(A - 2\chi)^{\frac{1}{2} } }
	\phi^{;\mu }\Phi_\mu \Phi ^\nu.
\end{align}
Looking at (\ref{dAlemshort}), we see that $\widehat \square \phi$ may be made proportional to $\square \phi$ if the second term on the right hand side of (\ref{ggphi}) vanishes. As before, this leads to the ``orthogonality'' condition given by $\phi^{;\mu }\Phi_\mu = 0$. 

While this condition is a significant stepping stone leading to the desired final form of $ C $ and $D$, it is not sufficient for the proportionality between $\widehat \square \phi$ and $\square \phi$; the disformal factor $ B $ should be related to $ A $. We have upon performing the second differentiation and dividing the result by $\sqrt{-\widehat g}$ to form $\widehat \square \phi$, gain
\begin{align}
	\widehat \square \phi
	=
	\frac{\square \phi}{A}
	+
	\frac{[A^{\frac{n-3}{2} }(A - 2\chi )^{\frac{1}{2}}]_{,\nu}
		g^{\mu\nu}\sqrt{-g}\,\phi_{;\mu }}
	{A^{\frac{n-1}{2}}(A - 2\chi)^{\frac{1}{2}}\sqrt{-g}}.
\end{align}
The second term vanishes for $A^{\frac{n-3}{2} }(A - 2\chi )^{\frac{1}{2} } = \bar c$, where $\bar c$ is some constant. This condition allows us to find the relationship between $B$ and $A$. Indeed, noting our ``orthogonality'' condition for $\phi^{;\mu}$ and $\Phi_\mu$, we get $\chi = -BX(Y^2 + 2XZ)/Y^2$. In effect,
\begin{align}
	\label{Beq2}
	B
	=
	-\frac{Y^2(A - \bar c^2A^{3 - n})}{2X(Y^2 + 2XZ)}
	\quad\text{and}\quad
	D^2
	=
	\frac{2X(bA^{3 - n}-A)}{Y^2 + 2XZ}
\end{align}
which matches the pair of equations given by (\ref{condsymkg}) in the immediately preceding section with the identification $\bar c^2 = b$.

The ``orthogonality'' condition binds $D$ and $C^2 = B$ in the general disformal transformation given by $\widehat g_{\mu\nu} = Ag_{\mu\nu} + (C\phi_{;\mu} + DX_{;\mu})(C\phi_{;\nu} + DX_{;\nu})$. On the other hand, the condition given by (\ref{Beq2}) above binds $B$ and $A$. In short, both $C$ and $D$ in the transformation are functionals of $A$ leaving only the conformal factor free. Provided that these conditions hold, the massless Klein-Gordon equation exhibits symmetry under the general disformal transformation---restating the same conclusion in the previous section.  For the massive case, the mass may be seen as transforming\footnote{This is quite an intriguing transformation. But as to its physical justification, making a `real-world' example of this process remains a part of our future studies.} from $m^2 \rightarrow \widehat m^2 = m^2/A$, in going from the manifold described by $g_{\mu\nu}$ to that with $\widehat g_{\mu\nu}$. 

\section{Invertibility of the Constrained General Disformal Transformation}
\label{secDTinvert}
The usual disformal transformation can be inverted in the sense that $g_{\mu\nu}$ can be solved using the hatted quantities. Indeed, given $\widehat g_{\mu\nu} = A(\phi,X)g_{\mu\nu} + B(\phi,X)\phi_{;\mu}\phi_{;\nu}$, one finds that $\widehat X = X/(A - 2BX)$, indicating that $X$ can be considered as a functional of $(\phi, \widehat X)$ provided\footnote{Given $\widehat X = X/(A - 2BX)$, the conformal and disformal factors, $A$ and $B$, respectively, must have restrictions on their functional forms in order to find the analytical equation for $X$ in terms of $\widehat X$. However, even if such an analytical equation does not exist, the inverse can still mathematically exist under the conditions stated in Refs.~\cite{Kobayashi:2019hrl,Bettoni:2013diz}.} $\partial \widehat X/\partial X \ne 0$; more precisely \cite{Kobayashi:2019hrl,Bettoni:2013diz}, $B(A - A_X X + 2B_X X^2) \ne 0$ and $ A - 2BX >0 $. As a consequence of this, we may simply rewrite the conformal and disformal factors $A$ and $B$ as a functional of $(\phi,\widehat X)$. In effect, the inverse disformal transformation may be written as
\begin{align}
	g_{\mu\nu}
	=
	\frac{\widehat g_{\mu\nu}}{A(\phi,\widehat X)}
	-
	\frac{B(\phi,\widehat X)}{A(\phi,\widehat X)}
	\phi_{;\mu}\phi_{;\nu};
\end{align}
or, in the same form as the usual disformal transformation, $g_{\mu\nu} = \widehat A(\phi,\widehat X)\widehat g_{\mu\nu} + \widehat B(\phi,\widehat X)\phi_{;\mu}\phi_{;\nu}$, with the identifications given by $\widehat A = 1/A$ and $\widehat B = -B/A$. 

The invertibility of disformal transformation is crucial in the study of actions describing scalar-tensor theories; i.e., Horndeski theory and beyond \cite{Horndeski:1974wa,Gleyzes:2014dya,Zumalacarregui:2013pma,Crisostomi:2016czh,BenAchour:2016fzp}. In the simplest scenario wherein both $A$ and $B$ are functionals of $\phi$ alone as in Ref.~\cite{Bettoni:2013diz}, it effectively maps the Horndeski theory back to itself; thus, preventing the existence of higher order derivative terms (in the action or equation of motion) signaling Ostrogradsky instability \cite{Ostrogradsky:1850fid,Woodard:2015zca}. Generalisations beyond this lead to beyond-Horndeski terms whose corresponding equations of motion go beyond second order. Nevertheless, if the transformation is invertible, the number of propagating physical degrees of freedom generally remains the same as in the original theory \cite{Domenech:2015tca,Takahashi:2017zgr,Arroja:2015wpa}.

The Klein-Gordon action is a `small' part but an important special case of the much more complicated Horndeski and beyond-Horndeski actions. However, it is important that we gain at least an insight about the invertibility of our general disformal transformation if the transformed action is to be physically sensible. To be honest, in the most general context, the transformation given by $\widehat g_{\mu\nu} = Ag_{\mu\nu} + (C\phi_{;\mu} + DX_{;\mu})(C\phi_{;\nu} + DX_{;\nu})$, with $(A, C, D)$ all functionals of $(\phi, X, Y, Z)$, is \textit{not} invertible; invertibility needs to be supplemented by some requirements on functional dependencies \cite{Takahashi:2021ttd}. In this section, our task is to find the invertibility conditions for this transformation subject to the forms of $C$ and $D$ given by (\ref{condsymkg}), ensuring the invariance of the massless Klein-Gordon equation. Furthermore, we wish to derive the inverse general disformal transformation to the extent permitted by this pair of equations.

As a requirement for the invertibility of the general disformal transformation (\ref{genDT}), we impose
that all the hatted terms $\widehat X, \widehat Y,$ and $\widehat Z$ be functionals of the unhatted quantities, namely, $(\phi, X, Y, Z)$, alone. With this restriction, we may invert the relationship between the hatted and unhatted quantities provided that the \textit{Jacobian determinant} is non-vanishing; i.e.,
\begin{align}
	\left|
	\begin{array}{ccc}
		\partial _X\widehat X 
		& 
		\partial _Y \widehat X 
		& 
		\partial_Z \widehat X
		\\[0.35em]
		\partial _X\widehat Y
		& 
		\partial _Y \widehat Y
		& 
		\partial_Z \widehat Y
		\\[0.35em]
		\partial _X\widehat Z
		& 
		\partial _Y \widehat Z
		& 
		\partial_Z \widehat Z							
	\end{array}
	\right|
	\ne 
	0.
\end{align}
For the case in hand, the restrictions among the functional factors in the general disformal transformation given by (\ref{condsymkg}) allow us to `spell out' the map $(\phi, X, Y, Z) \rightarrow (\widehat \phi,\widehat X, \widehat Y, \widehat Z) $; to a good extent, we can also determine the inverse map $(\widehat \phi,\widehat X, \widehat Y, \widehat Z)\rightarrow  (\phi, X, Y, Z)$. 

Starting with $\widehat X$ we find from its definition together with the equations for $\widehat g^{\mu\nu}$ and $ \Phi ^{\mu }\phi _{;\mu} = 0$,
\begin{align}
	\label{XoverA}
	\widehat X 
	=
	-\frac{1}{2A}\bigg(
	g^{\mu \nu } - \frac{\Phi ^\mu \Phi ^\nu }{A - 2\chi } 
	\bigg) 
	\phi _{;\mu }\phi _{;\nu }
	=
	\frac{X}{A}.
\end{align}

\noindent
It is, in other words, a functional of $ (\phi, X, Y, Z) $ only. For $\widehat Y$, we find an additional dependence on $\phi^{;\alpha}Y_{;\alpha}$ and $\phi^{;\alpha}Z_{;\alpha}$. Indeed,
\begin{align}
	\widehat Y
	&=
	\widehat g^{\mu \nu }\phi _{;\mu }\widehat X_{;\nu }
	=
	\frac{Y}{A^2} 
	+
	\frac{2A_\phi X^2}{A^3} 
	-
	\frac{A_XXY}{A^3} 
	-
	\frac{A_YX}{A^3} \phi ^{;\alpha } Y_{;\alpha }
	-
	\frac{A_ZX}{A^3} \phi ^{;\alpha } Z_{;\alpha }.
\end{align}
It follows that if $\widehat Y$ is to be dependent on $(\phi, X, Y, Z)$ alone, then the partial derivatives of the conformal factor, namely, $A_Y$ and $A_Z$ should both vanish leading to
\begin{align}
	\label{yhaty}
	\widehat Y
	&=
	Y\frac{A - A_XX}{A^3} 
	+
	\frac{2A_\phi X^2}{A^3}.
\end{align}
Following the same logic for $ \widehat Z $ we find
\begin{align}
	\label{zhatz}
	\widehat Z
	&=
	\frac{(A - A_X X)^2 (Y^2 + 2XZ)( {{A}^{n}} - bA^2) }{2bA^7X}
	\nonumber
	\\[0.5em]
	&\qquad
	+\,
	\frac{
		(A - A_X X)^2  Z
		-
		2A_\phi XY(A - {A_X}{X})
		-
		2 A_\phi^2 X^3
	}{A^5}.
\end{align}
It is worth noting that although $ A = A(\phi ,X) $ only, the disformal factors still depend on $ (\phi ,X,Y,Z) $ based on (\ref{condsymkg}).

In summary, assuming $A = A(\phi, X)$, we have a `ladderised' sequence of functional dependencies as
\begin{align}
	\label{eqnlad}
	\widehat X = \widehat X (\phi,X),
	\quad
	\widehat Y = \widehat Y(\phi,X,Y),
	\quad\text{and}\quad
	\widehat Z = \widehat Z(\phi,X,Y,Z).
\end{align}
Using (\ref{XoverA}), (\ref{yhaty}), and (\ref{metinvdt}) we may invert these relations as
\begin{align}
	\label{triohat}
	X &= A\widehat X,
	\quad
	Y
	=
	\frac{A\widehat Y - 2A_\phi \widehat X^2}
	{\widehat X_X},
	\nonumber
	\\[0.5em]
	Z
	&=
	\frac{b\widehat Z}{\widehat X_X^2 A^{n-3}} 		
	-
	\frac{(A\widehat Y - 2A_\phi \widehat X^2)^2( 1 - bA^{2-n}) }
	{2A\widehat X\widehat X_X^2}
	+
	\frac{
		2bA_\phi \widehat X(A\widehat Y - A_\phi \widehat X^2)
	}{\widehat X_X^2 A^{n-1}}.	
\end{align}
Note that the conformal factor, being unspecified, remains implicit in these equations. The existence of the full \textit{analytical} set of equations for the inverse relations depends on the form of $ A $. In the limit where $A = A(\phi)$ alone, the set of relations above completely solves for $(\phi,X,Y,Z)$ in terms of $(\phi, \widehat X, \widehat Y, \widehat Z)$. For general $ A = A(\phi ,X) $, barring the divergences of the equations above, the inverse relations exist, (although possibly, \textit{not} analytically,) provided $ (A -  A_XX)/A^2\ne 0 $ in accord with Ref.~\cite{Kobayashi:2019hrl}.

Having found $(\phi,X,Y,Z)$ in terms of $(\phi, \widehat X, \widehat Y, \widehat Z)$, we seek to find the inverse of the general disformal transformation; the aim is to write from 
\begin{align}
	\label{metinvdt}
	\widehat g_{\mu \nu }
	&=
	Ag_{\mu \nu }
	+
	(C\phi _{;\mu } + DX_{;\mu })
	(C\phi _{;\nu  } + DX_{;\nu  }),
	\nonumber
	\\[0.5em]
	\text{to}\quad
	g_{\mu \nu } 
	&= 
	\widehat A \widehat g_{\mu \nu } 
	+ 
	(\widehat C\phi _{;\mu } + \widehat D \widehat X_{;\mu })
	(\widehat C\phi _{;\nu } + \widehat D \widehat X_{;\nu }).	
\end{align}
In fulfilling this aim, we note from the first of (\ref{triohat}) that $ X_{;\mu}$ does not depend on the derivative term $\widehat X_{;\mu}$ alone but also includes $\phi_{;\mu}$. This leads to $\phi_{;\mu}X_{;\nu}$ producing $\phi_{;\mu}\phi_{;\nu}$ and $\phi_{;\mu}\widehat X_{;\nu}$; in particular, we have
\begin{align}
	\label{hattounhat}
	X_{;\mu }
	&=
	\frac{A_\phi \phi _{;\mu } \widehat X}
	{A\widehat X_X}
	+
	\frac{\widehat X_{;\mu }}
	{\widehat X_X}.    
\end{align}

At first sight, this might appear as a complication in deriving the second of (\ref{metinvdt}) due to additional hatted terms  `generated'. The fear is that we might \textit{not} end up with exactly the right form matching the original disformal transformation. However, as it turns out upon using (\ref{hattounhat}),
\begin{align}
	\label{boxinvdt}
	\boxed{
		g_{\mu \nu }
		=
		\frac{\widehat g_{\mu \nu }}{A} 	
		-
		\frac{AC^2}
		{(A\widehat Y - 2A_\phi \widehat X^2)^2}
		(\widehat Y \phi _{;\mu } + 2\widehat X\widehat X_{;\mu })
		(\widehat Y \phi _{;\nu } + 2\widehat X\widehat X_{;\nu }).
	}
\end{align}
Our sought-for inverse of the general disformal transformation, subject to the conditions given by (\ref{condsymkg}), takes the form consistent with that of the second of (\ref{metinvdt}). Moreover, it is worth noting that with the identifications of $\widehat C$ and $\widehat D$ in (\ref{boxinvdt}), we find $\widehat D \widehat Y = 2\widehat C\widehat X$, reflecting the same form as the original ``orthogonality'' condition.

\section{Concluding Remarks}
\label{secConclud}
Symmetry principles are ubiquitous in the fundamental laws of nature. In the  20th century, it dominated the exploration and formulation of theories in modern physics. With the realisation of its significance in understanding our Universe, we generalised a new class of symmetry related to the background metrical structure in which a field propagates. This new symmetry is induced by a broader class of metric transformations called the disformal transformation generalising conformal transformation that is extensively used in relativity and extensions thereof.

In this work, we extend the discovered symmetry under Bekenstein's disformal transformation of the massless Klein-Gordon equation. In particular, we show that, subject to some conditions, the massless relativistic wave equation for a scalar field in arbitrary dimensions is invariant under the general disformal transformation of the form given by $g_{\mu\nu} \rightarrow \widehat g_{\mu\nu} = A g_{\mu\nu} + \Phi_\mu \Phi_\nu$, where $\Phi_\mu = C\phi_{;\mu} + DX_{;\mu}$, with the functionals $A, C,$ and $D$ dependent on $X = -\tfrac{1}{2}\phi^{;\mu}\phi_{;\mu},\, Y = \phi^{;\mu}X_{;\mu}$, and $Z = X^{;\mu}X_{;\mu}$. The conditions involve a functional relationship between the disformal factors $(C,D)$ and the conformal factor $A$. In addition to this, with an insight for the preservation of the number of physical degrees of freedom upon transformation, we derive the inverse disformal transformation. The inverse may exist if the conformal factor is restricted to $A = A(\phi,X)$. Interestingly, while disformal factors $C$ and $D$ are related to $A$ in view of the symmetry of the Klein-Gordon equation, their functional dependencies include $(Y,Z)$ in addition to $(\phi,X)$ on which the conformal factor depends.

For future studies, we have at least three tracks in sight. First, we can revisit the other fundamental equations in physics, e.g., the Dirac equation and Maxwell's equations, that may exhibit an extended symmetry under a hierarchy of disformal transformations. Second, we may ponder on the mass transformation, together with the interaction terms in the action, induced by disformal transformation. Lastly, it would be interesting to investigate disformal transformation, at least of the scalar fields and derivatives thereof, within the realm of quantum field theory in curved spacetime.

\section{Acknowledgment}
MRD Chua would like to thank the Department of Science and Technology Science Education Institute (DOST-SEI) for providing financial assistance in support of this study.

\section{Data availability statement}
No new data were created or analysed in this study.

\section{ORCID iDs}
Allan L. Alinea\; \url{https://orcid.org/0000-0001-7511-1256}\\
Mark Ruel D. Chua\; \url{https://orcid.org/0000-0002-2451-0238}

\end{document}